\documentclass[aps,twocolumn,showpacs,preprintnumbers,amsmath,amssymb,prl]{revtex4-1}
\usepackage{epsf,amsmath,amssymb,verbatim,color,multirow,pifont}
\usepackage{graphicx}

\begin{document}
\newcommand{\tx}{{t_{\rm x}}}

\title{Suppression effect on explosive percolations}
\author{Y.S. Cho and B. Kahng}
\affiliation{{Department of Physics and Astronomy, Seoul National University,
Seoul 151-747, Korea}}
\date{\today}

\begin{abstract}
Percolation transitions (PTs) of networks, leading to the formation of a macroscopic cluster, are conventionally 
considered to be continuous transitions. However, a modified version of the classical random graph 
model was introduced in which the growth of clusters was suppressed, and a PT occurs explosively 
at a delayed transition point. Whether the explosive PT is indeed discontinuous or continuous 
becomes controversial. Here, we show that the behavior of the explosive PT depends on 
detailed dynamic rules. Thus, when dynamic rules are
designed to suppress the growth of all clusters, the discontinuity
of the order parameter tends to a finite value as the system size increases,
indicating that the explosive PT could be discontinuous.
\end{abstract}

 \maketitle
Percolation transition (PT), i. e., the transition from a disconnected state to a connected one, has been
regarded as a fundamental model of phase transitions in nonequilibrium systems~\cite{percolation}.
The concept of PT has been extended to the formation of macroscopic clusters
in network science. A pioneering model of  PT in network science is the classical
random graph model introduced by Erd\H{o}s and R\'enyi (ER)~\cite{er} in which a system composed
of a fixed number of vertices $N$ evolves as edges are added. At each evolution step,
an edge is added between two vertices, that are randomly selected from among
unconnected pairs of vertices. In this model, a quantity, called time, is defined as the number
of edges added to the system per node. 
The ER model has been modified by following the so-called Achlioptas
process~\cite{science}.  The Achlioptas process essentially identifies the dynamics
that prevent the creation of a given target pattern by choosing one edge from a given number 
of randomly selected potential edges. 
In the modified ER model for the PT, the target pattern is a giant cluster. 
Thus, an edge to be added to the system should be selected such that the growth 
of clusters can be systematically suppressed. The principle to take this selection rule 
is hereafter referred  to as the suppression principle (SP).

The dynamic rule originally designed by Achlioptas et al.~\cite{science} is as follows:
in the case of two randomly selected edge candidates, the one actually added to the 
system is the one minimizing the product or the sum of the sizes of the clusters 
that are connected by each potential edge. 
The ER models modified according to the product rule and the sum rule are 
denoted as ERPR and ERSR models, respectively. 
In these models, the giant-cluster size increases drastically at the critical point, 
and therefore, the percolation transition is called explosive percolation. 
The introduction of an explosive PT model has triggered intensive
research on discontinuous PTs in non-equilibrium
systems~\cite{ziff,cho_sf,santo,friedman,herrmann1,can,cho_fss,souza,araujo}.
Many models have followed the ERPR and ERSR models,
and they display similar transition patterns. Although the explosive PT was regarded
as discontinuous in the original paper~\cite{science}, recently it has been argued that the
transition is continuous in the thermodynamic limit~\cite{mendes,paczuski,andrade,hklee,warnke}.
Thus, the issue of whether the explosive PT is indeed discontinuous or continuous remains
controversial. In this Letter, we describe the microscopic investigation of the dynamic rules of several explosive
percolation models and the surprising finding that these rules do not satisfy
the SP. Thus, it is rather natural that the PTs of those models are continuous.
However, some other variants of the Achlioptas model satisfying the SP
exhibit the pattern of discontinuous PTs within the range of our numerical simulations. 
Thus, we may state that satisfying the SP is essential for discontinuous PTs in the evolution of 
complex networks.

\begin{figure}[t]
\includegraphics[width=1\linewidth]{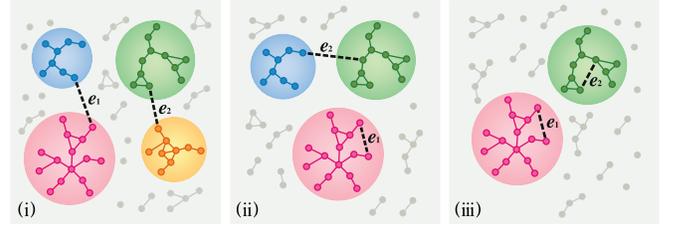}
\caption{(Color online) Classification of types of edge candidate pairs.
(i) Both candidates $e_{1}$ and $e_{2}$ are intercluster edges.
(ii) $e_{1}$ is an intracluster edge and the other candidate $e_{2}$ is an intercluster edge.
(iii) Both $e_{1}$ and $e_{2}$ are intracluster edges.}
\label{fig:cases}
\end{figure}

To explain the dynamic rule, we classify the types of edge candidate pairs as follows:
\begin{itemize}
\item[(i)] Both edge candidates $e_1$ and $e_2$ are intercluster edges. Clusters
of sizes $s_{1a}^{\rm (i)}$ and $s_{1b}^{\rm (i)}$ are connected by the edge $e_1$, and
clusters of sizes $s_{2a}^{\rm (i)}$ and $s_{2b}^{\rm (i)}$ are connected by the edge $e_2$.
In the following discussion, we will use the following notation: $P_1^{\rm (i)}=s_{1a}^{\rm (i)}s_{1b}^{\rm (i)}$,
$P_2^{\rm (i)}=s_{2a}^{\rm (i)}s_{2b}^{\rm (i)}$,
$S_1^{\rm (i)}=s_{1a}^{\rm (i)}+s_{1b}^{\rm (i)}$ and
$S_2^{\rm (i)}=s_{2a}^{\rm (i)}+s_{2b}^{\rm (i)}$.
\item[(ii)] $e_1$ is an intracluster edge in a cluster of size $s_1^{\rm (ii)}$,
and $e_2$ is an intercluster edge between two clusters of sizes
$s_{2a}^{\rm (ii)}$ and $s_{2b}^{\rm (ii)}$. We denote
$P_2^{\rm (ii)}=s_{2a}^{\rm (ii)}s_{2b}^{\rm (ii)}$ and
$S_2^{\rm (ii)}=s_{2a}^{\rm (ii)}+s_{2b}^{\rm (ii)}$.
\item[(iii)] Both $e_1$ and $e_2$ are intracluster edges in either
the same cluster or two different clusters.
\end{itemize}
These three types of edge candidate pairs are depicted in Fig.~\ref{fig:cases}.
On the basis of this classification, we formulated several dynamic rules
to determine which edge should be added to the system.

\begin{figure}[t]
\includegraphics[width=0.9\linewidth]{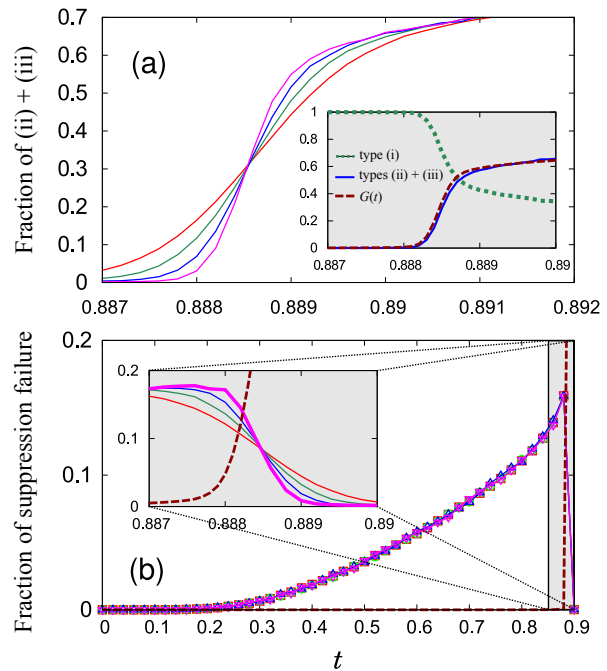}
\caption{(Color online) (a) The fractions of type (ii) and type (iii) potential edges as a
function of time $t$ for different system sizes: $N/10^4 = 32, 64, 128$, and 256 from
the top (bottom) in the small-$t$ (large-$t$) region. As $N$ increases, the fraction
increases dramatically. Inset: For a given $N = 1.024 \times 10^7$, the
fractions of the type (i) (dotted line) and the types (ii) and (iii) (solid lines)
as a function of $t$. They are compared with the giant-cluster size $G_{N}(t)$ (dashed line).
The fraction of the types (ii) or (iii) behaves similarly to $G_{N}(t)$, indicating that 
counting for the effect by taking the intracluster edges becomes important as $t$ 
approaches $t_c$.
(b) The fraction of the occurrences in which the sum of the sizes of one pair of clusters
becomes larger than that of the size of the other pair, even though the product of the sizes 
of the former is smaller than that of the sizes of the latter.
The dotted line represents $G_N(t)$.
Inset: Solid lines represent the failure ratio on an enlarged scale
for different system sizes $N/10^{4} = 32, 64, 128$, and 256.
For a larger system, the curve lies on the upper position in the small-$t$
region. The dashed line represents $G_N(t)$. Numerical data for (a) and (b) are 
obtained from the ERPR-B model.}
\label{fraction}
\end{figure}

Although the original model~\cite{science} seems to follow the basic
idea of the Achlioptas process, the dynamic rule has to be more carefully examined.
As time approaches the percolation threshold, the mean cluster size increases, and 
one or both potential edges have a greater possibility of being intracluster as shown in
Fig.~\ref{fraction}(a). Thus, we have to clarify how to formulate the dynamic rule
when intracluster edges are selected as potential edges.
Here, we formulate dynamic rules for the cases (i)-(iii), and we check whether each rule
does follow the SP.

First, we introduce three different variants of the ERPR model; these
are specified in Table I. In model A, when one edge is an intracluster edge
((ii) and (iii)), the product is the square of the size of that cluster,
while for case (i), it is the product of the sizes of the two clusters connected by one intercluster
edge. This rule, however, can fail to follow the Achlioptas
SP. For example, when $s_1^{\rm (ii)}=5$, $s_{2a}^{\rm (ii)}=3$, and
$s_{2b}^{\rm (ii)}=7$ for case (ii), the edge $e_2$ is selected in the ERPR-A 
model, and then, the size of the created cluster is 10.
On the other hand, if edge $e_1$ is selected, then none of the clusters would increase in size.
Therefore, model A does not follow the SP.
As the transition point is approached, intracluster edges in (ii) and (iii)
can be selected more frequently [see Fig.~\ref{fraction}(a)]. Actually, the
behavior of the fraction of the occurrences of (ii) or (iii) is similar to that of 
$G(t)$. Thus the failure of the SP can be more frequent.

In the ERPR-B model, when potential edges of the type (ii) are selected, the intracluster edge
is definitely selected, so that the cluster size does not increase. When the two potential
edges are both intracluster edges (type (iii)), one of them is randomly selected. By this rule,
time is advanced by one unit $1/N$ for the types (ii) and (iii).

\begin{table*}[t]
\caption{List of the dynamic rules under the product rule [the sum rule]. The second column lists
the type of potential edges. The third and fourth columns list the conditions for the cases
in which both $e_1$ and $e_2$ in Fig. 1 are selected.
The last column shows the case when either $e_1$ or $e_2$ is selected randomly.}
\begin{ruledtabular}
\begin{tabular}{ccccc}
Model & Type & $e_1$ & $e_2$ & {\rm Either $e_1$ or $e_2$ randomly}\\
\hline
\begin{tabular}{c}
Model A
\end{tabular} &
\begin{tabular}{c}
Type (i) \\
Type (ii) \\
Type (iii)
\end{tabular} &
\begin{tabular}{c}
$P_1^{\rm (i)} < P_2^{\rm (i)}$ [$S_1^{\rm (i)} < S_2^{\rm (i)}$] \\
$(s_1^{\rm (ii)})^2 < P_2^{\rm (ii)}$ [$ 2s_1^{\rm (ii)} < S_2^{\rm (ii)}$] \\
~~
\end{tabular} &
\begin{tabular}{c}
$P_1^{\rm (i)} > P_2^{\rm (i)}$ [$S_1^{\rm (i)} > S_2^{\rm (i)}$] \\
$(s_1^{\rm (ii)})^2 > P_2^{\rm (ii)}$ [$2s_1^{\rm (ii)} > S_2^{\rm (ii)}$] \\
~~
\end{tabular} &
\begin{tabular}{c}
$P_1^{\rm (i)} = P_2^{\rm (i)}$ [$S_1^{\rm (i)} = S_2^{\rm (i)}$]\\
$(s_1^{\rm (ii)})^2 = P_2^{\rm (ii)}$ [$2s_1^{\rm (ii)} = S_2^{\rm (ii)}$] \\
unconditional
\end{tabular} \\
\\
\begin{tabular}{c}
Model B
\end{tabular} &
\begin{tabular}{c}
Type (i) \\
Type (ii) \\
Type (iii)
\end{tabular} &
\begin{tabular}{c}
$P_1^{\rm (i)} < P_2^{\rm (i)}$ [$S_1^{\rm (i)} < S_2^{\rm (i)}$]\\
unconditional \\
~~
\end{tabular} &
\begin{tabular}{c}
$P_1^{\rm (i)} > P_2^{\rm (i)}$ [$S_1^{\rm (i)} > S_2^{\rm (i)}$]\\
~~ \\
~~
\end{tabular} &
\begin{tabular}{c}
$P_1^{\rm (i)} = P_2^{\rm (i)}$ [$S_1^{\rm (i)} = S_2^{\rm (i)}$]\\
~~\\
unconditional
\end{tabular} \\
\\
\begin{tabular}{c}
Model C
\end{tabular} &
\begin{tabular}{c}
Type (i) \\
\end{tabular} &
\begin{tabular}{c}
$P_1^{\rm (i)} < P_2^{\rm (i)}$ [$S_1^{\rm (i)} < S_2^{\rm (i)}$]\\
\end{tabular} &
\begin{tabular}{c}
$P_1^{\rm (i)} > P_2^{\rm (i)}$ [$S_1^{\rm (i)} > S_2^{\rm (i)}$]\\
\end{tabular} &
\begin{tabular}{c}
$P_1^{\rm (i)} = P_2^{\rm (i)}$ [$S_1^{\rm (i)} = S_2^{\rm (i)}$]\\
\end{tabular}
\end{tabular}
\end{ruledtabular}
\end{table*}

\begin{figure}
\includegraphics[width=1.0\linewidth]{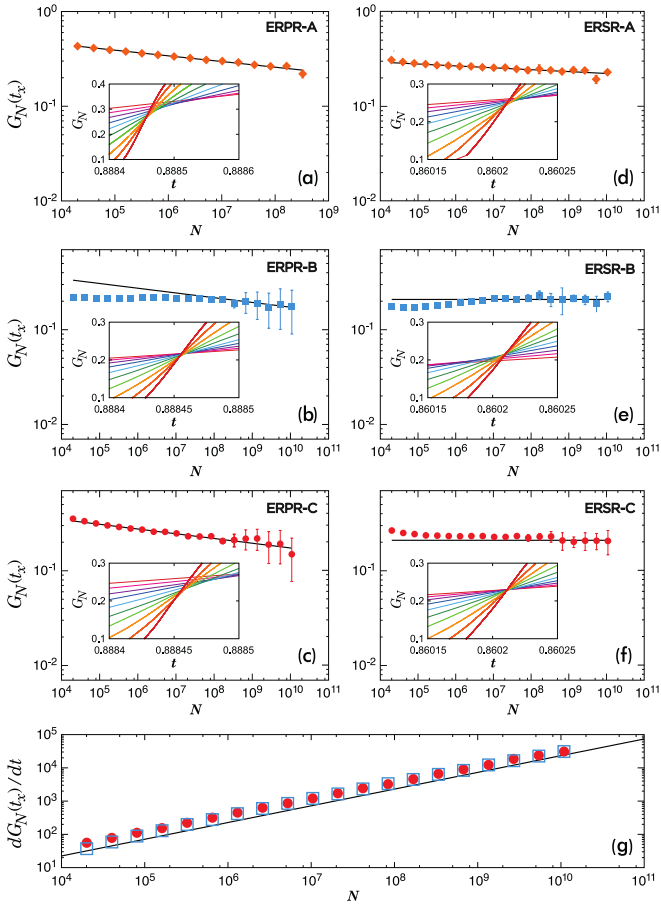}
\caption{(Color online) $G_{N}(\tx)$ versus $N$ for the (a) ERPR-A, (b) ERPR-B,
(c) ERPR-C, (d) ERSR-A, (e) ERSR-B, and (f) ERSR-C models.
The slopes of each guideline is (a) $-0.06$, (b) $-0.05$, (c) $-0.05$,
(d) $-0.02$, (e) $0$, and (f) $0$.
Thus, $G_N(\tx)$ of ERSR-B and ERSR-C converge to finite values
in the thermodynamic limit. The error bars represent the deviation of the
cross points. Each data set is obtained by taking an average over
about $10^{13}/N$ configurations. $G_N(\tx)$ of models B and C of
both the ERPR and ERSR overlap in the large-$N$ region.
The insets of (a)-(f) show the behaviors of $G_{N}(t)$ for different system
sizes $N/10^{4} = 32, 64, 128, 256, 512, 1024, 2048, 4096, 8192$, and 16384.
(g) Plot of the slope of $G_N(t)$ at $\tx$ versus $N$ for
the ERSR-B ($\square$) and the ERSR-C ($\circ$) models.
$dG_N(\tx)/dt$ increases according to a power law $\sim N^{0.5}$.
}
\label{fig3}
\end{figure}

\begin{figure}
\includegraphics[width=1.0\linewidth]{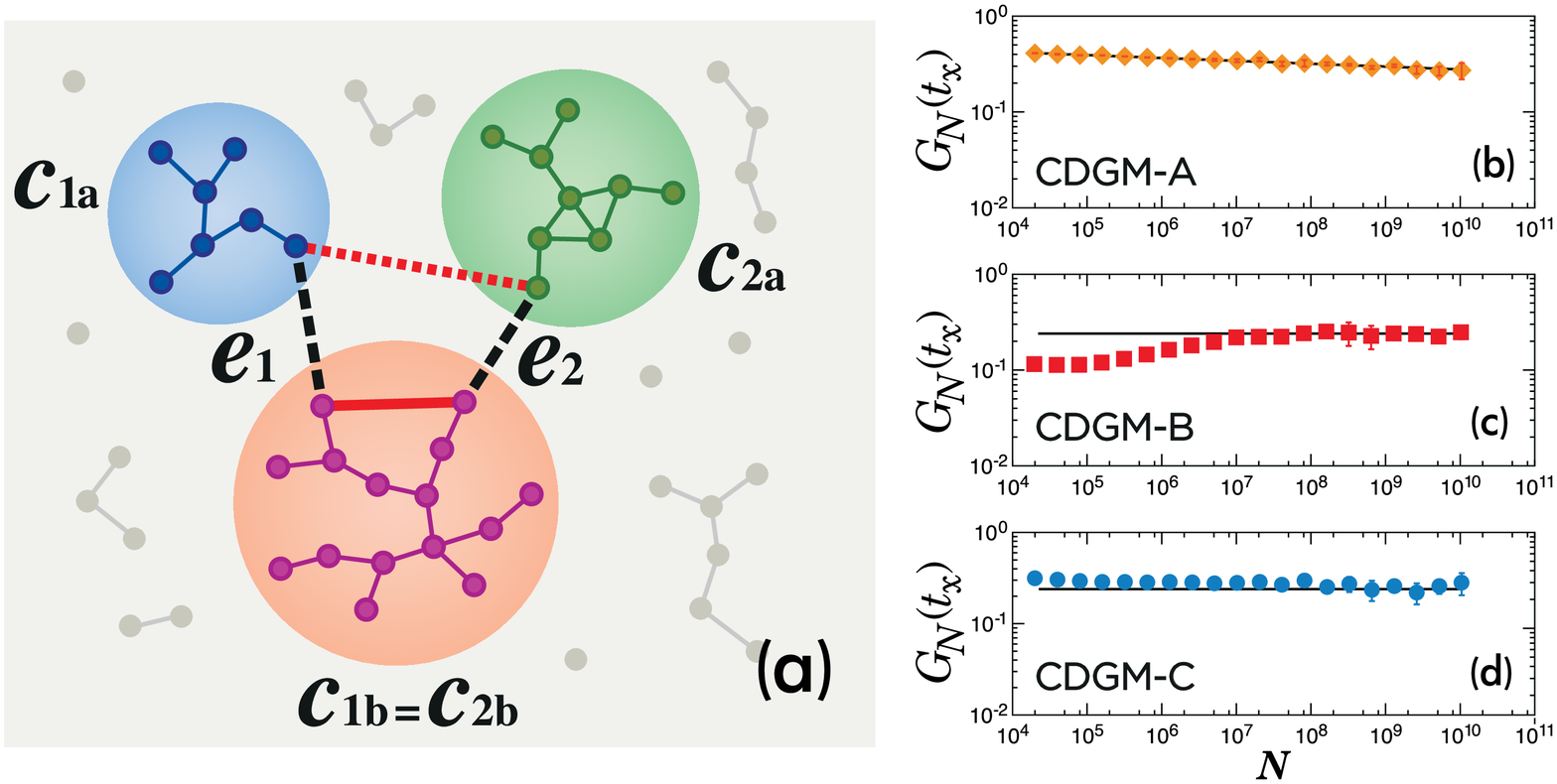}
\caption{(Color online) (a) Schematic illustration of the selection rule on the 
intracluster edge for the CDGM-B model. In the original CDGM model (CDGM-A), 
when two pairs of edges $e_1$ and $e_2$ are selected from the
cluster sets $C_1$ and $C_2$ of sizes $s_{1a} < s_{1b}$ and $s_{2a} < s_{2b}$,
an additional edge connects the two nodes in the clusters $C_{1a}$ and $C_{2a}$
(dotted line). In our modified model (CDMA-B) the two nodes in the
clusters $C_{1b}$ and $C_{2b}$ are connected instead (solid line),
because by choosing them, the cluster sizes in the system
do not increase.
Panels (b)-(d) show $G_N(\tx)$ versus $N$ for CDGM-A, CDGM-B, and CDGM-C models, respectively, and 
the slopes are $-0.03$, 0 and 0 respectively. $G_N(\tx)$ of the models B and C
overlap in the large $N$ region.
It may be reasonable to expect that the modified CDGM model (CDGM-B and C) shows a discontinuous PT. }
\label{fig4}
\end{figure}

Model C is a simplified version of models A and B. In this model,
the dynamics proceed via only intercluster connections.
Thus, two potential edges are both intercluster edges (case (i)).
Model C may be regarded to be nearly the same as model B because the clusters do not
grow when the intracluster edge is selected in (ii) and (iii).
However, the difference between them is that for types (ii) and (iii), time is
advanced in model B but not in model C. 

For all models A, B, and C, the product rule has an intrinsic drawback that
the Achlioptas SP is unfulfilled.
Let us consider a simple example of two intercluster connections, in which
$s_{1a}^{\rm (i)}=2$, $s_{1b}^{\rm (i)}=7$, $s_{2a}^{\rm (i)}=4$,
and $s_{2b}^{\rm (i)}=4$. Then, $P_1^{\rm (i)}=14$ and $P_2^{\rm (i)}=16$,
and thus, edge $e_1$ is added to the system.
However, the resulting cluster size is 9 in the case, which is larger
than the resulting size 8 when $e_2$ is added. In other words, even though the 
product of one pair of cluster sizes is smaller than that of the other pair, its 
sum can be larger. Thus, the Achlioptas SP is inherently unfulfilled.
Investigations~\cite{mendes, hklee, cho_sf, cho_fss} have shown that the cluster
size distribution displays a hump shape in the region of large cluster sizes, and
the hump size increases up to the point where explosive cluster aggregations
start. Owing to the inherent drawback, such a case is likely to occur 
frequently, as shown in Fig.~\ref{fraction}(b). Thus, the PTs under the product 
rule are continuous regardless of the model type.

Next,  we introduce similar models under the sum rule; these are also listed in Table I.
The drawback inherent to the product rule is
removed in the ERSR model. However, for case (ii), the Achlioptas
SP cannot be fulfilled in model A, but it is always
fulfilled in models B and C. Thus,in the case of the sum rule, the models B and C  are regarded
to be the ones following the Achlioptas SP. We state that the PTs
for models B and C are possible candidates for the discontinuous PT.

Even though it is a challenging task to determine the transition types of the explosive
PTs with numerical simulation data, the numerical approach is the only one possible,
since there is no analytic solution that takes into account all the aforementioned cases.
Extensive numerical simulations are carried out up to a system size $N=10^{10}$
with a configuration average of about $10^{13}/N$. 
The obtained numerical data may be sufficient for understanding why this
controversy has arisen, but a higher configuration average may be required for
determining the type of PT, particularly when the system size is large.

We measure $G_N(t)$ as a function of time $t$ for different system sizes
in the range $N=2^0\times 10^4 - 2^{20}\times 10^4$ at every $N=2\times 10^4$ step.
For given $N$ and $2N$, we find a point of intersection of the two curves
$G_N(t)$ and $G_{2N}(t)$. The time and $G$ components of such a point are
denoted as $t_{\rm x}(2N)$ and $G_{2N}(\tx(2N))$, respectively.
We compose a set of $\{\tx(N), G_N(\tx(N)) \}$ for the simulated system sizes.
To evaluate the discontinuity of the PT, we propose the following criteria:
{\it
\begin{itemize}
\item[($\alpha$)] The value $(\tx(N), G_N(\tx))$ remains finite as
$N\to \infty$. The time $\tx(\infty)$ is regarded as the
transition point $t_c$ in the thermodynamic limit.
\item[($\beta$)] The tangent of the curve $G_N(t)$ with respect to
$t$ at $\tx(N)$ diverges as $N$ increases.
\end{itemize}}

Figs.~\ref{fig3}(a)-(c) show the behaviors of $G_N(\tx)$ as a function of $N$
for models A, B, and C, respectively, under the product rule.
The insets of each figure show $G_N(t)$ versus $t$.
We can see in (a) and (c) that for the model A and C under the product
rule, $G_N(\tx)$ decreases with increasing $N$ in the whole considered range, 
suggesting that $G_N(t_c)\to 0$ in the limit $N\to \infty$.
In the case of model B, even though the data look flat up to $N \simeq 10^8$,
they decay in the large-$N$ region in the same manner
as for model C. Thus, the decay behavior can be considered to stem from the intrinsic
drawback of the product rule for the case (i). In our previous study~\cite{cho_fss},
we performed numerical simulations for the model B under the product rule up to
a system size $N=10^8$. In that case, however, the decay behavior was not noticed and
the PT was regarded as a discontinuous transition. With the simulation
data obtained in this sutdy for larger system sizes, we conclude that
the three models based on the product rule show continuous PTs.

Figs. \ref{fig3}(d)-(f) are the plots of $G_{N}(t_{\rm x})$ versus $N$
for models A, B, and C under the sum rule.
It can easily be seen that for model A, the $G_N(t_{\rm x})$ values 
decrease with increasing $N$, suggesting that $G_N(t_c)$
is zero in the limit $N\to \infty$.
This indicates that the rule of doubling the cluster size in the
ERSR-A model violates the SP, and leads to a continuous transition.
However, for models B and C, the data of  the $G_N(t_{\rm x})$ values 
look relatively flat asymptotically within the large-$N$ limit, even though 
the data points beyond $N=10^9$ have large error bars owing to a smaller 
number of configuration averages.
Fig.~\ref{fig3}(g) shows the tangent to the curve $G_N(t)$ at the crossing point
as a function of $N$. The data show that the tangent increases according to a power law
$\sim N^{0.5}$. Thus, we may conclude that the ERSR-B and ERSR-C models
fulfill the SP and seem to show discontinuous PTs, within the range of our
numerical data. However, we cannot rule out the possibility that the value
of $G_N(\tx(N))$ decreases when the system sizes are larger than those  simulated in this work.

Recently, the authors \cite{mendes} introduced a new type of Achlioptas percolation 
model and argued that the explosive PT is actually continuous.
This model, called the CDGM model following the initials of the authors, is 
defined as follows:  Firstly, a pair of clusters, $C_{1a}$ and $C_{1b}$, are randomly selected  
and the smaller cluster (say $C_{1a}$) is selected. 
Secondly, another pair of clusters  $C_{2a}$ and $C_{2b}$  are randomly 
picked, and the smaller one (say $C_{2a}$) is selected. 
Thirdly, two random nodes from each of the chosen  
clusters ($C_{1a}$ and $C_{2a}$) are selected and  
connected. There are four possible combinations of the connection.
However, when either of the two clusters from the first set is identical 
to either of the two clusters from the second set (for example, 
$C_{1b}$ and $C_{2b}$ in Fig.~\ref{fig4}(a)), the CDGM model 
can fail to follow the Achlioptas SP.  According to the original 
rule of the CDGM model, the two smaller clusters $C_{1a}$ and 
$C_{2a}$ are connected, which creates a larger cluster size 
whereas the connection between two nodes inside the cluster $C_{1b}$ = 
 $C_{2b}$ does not increase any cluster size in the system. 
Therefore,the original CDGM model fails to follow the Achlioptas SP 
by not taking into account the natural choice of that intra-cluster edge.
Again, as time approaches $t_c$, the selection of an intracluster edge
becomes more frequent. Thus, the selection of intercluster edges, which 
is against the SP, can change the PT into a continuous transition. 
We confirm this result by performing the following numerical simulations.
We plot $G_N(\tx)$ versus $N$, and find $G_N(t_{\rm x}) \sim N^{-0.03}$
for the CDGM model (Fig.~\ref{fig4}(b)). We then modify the original model as 
follows: when one or two intracluster edges are present among the four edge candidates, 
one of these intracluster candidates is connected (model B) randomly.
In addition, we consider a model C in which the four edge candidates are only allowed 
to be intercluster edges. Figs.~\ref{fig4}(c) and (d) suggests that $G_N(\tx)$ approaches 
a constant value in the large-$N$ region for models B and C, respectively.
Moreover, we confirm that the slope of $G_N(t)$ at $\tx$ diverges
as $N$ increases in a power law manner. Based on these numerical results, 
it reveals that the PT could be discontinuous even in the CDGM model, when 
properly modified.

In summary, we have examined the dynamic rule of the Achlioptas model in the perspective
of the suppression principle (SP). We found that when the dynamic rule does
follow the SP, the numerically calculated order parameter seems to show the behavior 
of a discontinuous transition. Otherwise, the PT seems to be continuous. 
The original  Achlioptas model and the CDGM model belong to the latter case. 
An analytic study of the modified Achlioptas models by taking 
into account the suppression effect is therefore needed. 

This study was supported by an NRF grant awarded through the Acceleration
Research Program (Grant No. 2010-0015066) and the NAP of KRCF (BK) and the Seoul
Science Foundation (YSC).

\end{document}